 \documentclass[10pt,preprint]{aastex}  
 \usepackage{natbib}
\def\gapprox{\lower.4ex\hbox{$\;\buildrel >\over{\scriptstyle\sim}\;$}}
\def\lapprox{\lower.4ex\hbox{$\;\buildrel <\over{\scriptstyle\sim}\;$}}

\shortauthors{ASCHWANDEN}
\shorttitle{Self-Organized Criticality}

\begin{document}

\title{		Self-Organized Criticality in Solar and Stellar Flares: 
		Are Extreme Events Scale-Free ?} 

\author{        Markus J. Aschwanden$^1$}
\affil{         $^1)$ Lockheed Martin,
                Solar and Astrophysics Laboratory,
                Org. A021S, Bldg.~252, 3251 Hanover St.,
                Palo Alto, CA 94304, USA;
                e-mail: aschwanden@lmsal.com }

\begin{abstract}
We search for outliers in extreme events of statistical
size distributions of astrophysical data sets, motivated
by the {\sl Dragon-King hypothesis} of Sornette (2009),
which suggests that the most extreme events in a statistical
distribution may belong to a different population, and thus
may be generated by a different phyiscal mechanism, in contrast 
to the strict power law behavior of {\sl self-organized
criticality (SOC)} models. Identifying such disparate outliers 
is important for space weather predictions. Possible physical
mechanisms to produce such outliers could be generated by
sympathetic flaring. However, we find that Dragon-King events
are not common in solar and stellar flares, identified in
4 out of 25 solar and stellar flare data sets only. 
Consequently, small, large, and extreme flares are essentially
scale-free and can be modeled with a single physical mechanism.
In very large data sets ($N \gapprox 10^4$) we find significant
deviations from ideal power laws in almost all data sets.
Neverthess, the fitted power law slopes constrain 
physcial scaling laws in terms of flare areas and volumes,
which have the highest nonlinearity in their scaling laws.
\end{abstract}

\keywords{Sun: corona --- Sun: flares ---  Sun: X-rays, gamma rays  
      	 --- stars: flare --- methods: statistical }

\section{	INTRODUCTION				}

The largest natural catastrophes that can happen 
are obviously of highest interest, because they
cause the largest threats and damages, such as 
the biggest earth quakes, land slides, wild fires, 
volcanic eruptions, terrorist attacks, stock market 
crashes, solar flares, coronal mass ejections, 
or magnetospheric storms 
(see reviews and textbooks by Bak 1996; Jensen 1998; 
Charbonneau et al.~2001; Hergarten 2002; Sornette 2004; 
Aschwanden 2011, 2013; Aschwanden et al.~2016; Pruessner 2012; 
Watkins et al.~2016; Sharma et al.~2016).
The ubiquitous power law-like size distributions that are
universally found in the statistics of event sizes have
been attributed to nonlinear scale-free processes. 
The observed size distributions can generally be subdivided 
into three regimes: 
(i) a range of incomplete sampling below the inertial range, 
(ii) an inertial range over which the power law function holds, 
and (iii) a cutoff above the inertial range that approaches
effects of the finite system size. The three regimes can be 
associated with (i) small events, (ii) large events, 
and (iii) extreme events.
Generally, it is tacitly assumed that small, large, and
extreme events belong to the same population, the same
category of events, and be driven by the same physical 
event generation mechanism (for a given phenomenon).
However, it has recently been suggested that the most
extreme events do not develop similarly to the other 
events, at least for the case of stock market crashes,
which are predictable to some extent based on their
precursor behavior, a phenomenon that has been dubbed
the {\sl ``Dragon-King hypothesis''} (Sornette 2009;
Sornette and Ouillon 2012). This hypothesis suggests
that extreme events may result from amplification
and synchronization processes, and that the 
{\sl ``Big events''} do have characteristic fingerprints
 enabling their forecasting and prediction. 
Another application of this hypothesis is the sub-critical
failure of materials under load, where a long phase of
damage accumulation characterized by power law distributed
microfracture events finally give way to catastrophic
breakdown (Baro et al.~2013). A third application are
magnetospheric storms, where a growing recognition emerges
that long-range correlations are an essential feature of
systems that exhibit extreme events (Sharma 2017). 
In this study we investigate the question whether extreme 
events of solar and stellar flares are consistent with a power law 
distribution as an extension of a scale-free inertial range, 
or if they show significant deviations from ideal power laws.

\section{	DATA ANALYSIS AND RESULTS 		}

\subsection{ 	Definition of Power Law Distributions 	}	

An ideal power law distribution that represents a {\sl
differential occurrence frequency distribution}, also
simply called a {\sl size distribution} $N(x)$ of
events, as a function of some size parameter $x$, can be 
described by four parameters ($x_1, x_2, n_0, \alpha_x$), 
\begin{equation}
        N(x) dx = n_0 \ x^{-\alpha_x} dx \ ,
	\qquad x_1 \le x \le x_2 \ ,
\end{equation}
where $x_1$ and $x_2$ are the lower and upper bounds of
the power law inertial range, $\alpha_x$ is the power law 
slope, and $n_0$ is the normalization constant,
\begin{equation}
        n_0     =n_{ev} (1-\alpha_x)
        \left[ (x_2)^{1-\alpha_x}-(x_1)^{1-\alpha_x}
        \right]^{-1} \ .
\end{equation}
with $n_{ev}$ being the total number of events contained
in the inertial range $[x_1 \le x \le x_2]$.

The ideal power law function (Eq.~1) can be generalized
with an additional ``shift'' parameter $x_0$ (with respect to $x$), 
also called {\sl Lomax distribution} (Lomax 1954),
{\sl Generalized Pareto distribution} (Hosking and Wallis
1987), or {\sl Thresholded power law size distribution}
(Aschwanden 2015),
\begin{equation}
        N(x) dx = n_0 \left( x_0 + x \right)^{-\alpha_x} dx \ ,
\end{equation}
with the normalization constant $n_0$ for the range $x_1 \le x \le x_2$,
\begin{equation}
        n_0     =n_{ev} (1-\alpha_x)
                \left[ (x_2+x_0)^{1-\alpha_x}-(x_1+x_0)^{1-\alpha_x}
                \right]^{-1} \ .
\end{equation}
This additional parameter $x_0$ accomodates three different
features: truncation effects due to incomplete sampling of
events below some threshold (if $x_0 > 0$), incomplete
sampling due to instrumental sensitivity limits (if $x_0 > 0$),
or subtraction of event-unrelated background (if $x_0 < 0$),
as it is common in astrophysical data sets.
  
While Eqs.~(1) and (3) represent {\sl differential size
distributions} $N_{diff}(x)$, it is statistidally more
advantageous to employ {\sl cumulative size distributions} $N_{cum}(x)$, 
especially for small data sets and near the upper cutoff,
where we deal with a small number of events per bin. 
We will use cumulative size distributions that include all events
accumulated above some size $x$, such as for the {\sl ideal power law
function},
\begin{equation}
        N_{cum}(>x) dx = \int_{x}^{x_2} n_0\ x^{-\alpha_x} dx
        = 1 + (n_{ev}-1) \left( { x_2^{1-\alpha}-x^{1-\alpha} \over
                          x_2^{1-\alpha}-{x_1}^{1-\alpha} } \right) \ .
\end{equation}
or for the {\sl thresholded power law size distribution},
\begin{equation}
        N_{cum}(>x) = \int_{x}^{x_2} n_0
        ( x + x_0 )^{-\alpha_x} dx
        = 1 + (n_{ev}-1) \left( {(x_2+x_0)^{1-\alpha_x}-(x  +x_0)^{1-\alpha_x} \over 
                         (x_2+x_0)^{1-\alpha_x}-(x_1+x_0)^{1-\alpha_x} } \right) \ .
\end{equation}
We show an example of a size distribution that contains the counts
of events per bin $n_{cts}(x)$ in Fig.~(1a), the corresponding
differential size distribution $N_{diff}(x)=n_{cts}(x)/\Delta x$ in Fig.~(1b),
and the corresponding cumulative size distribution 
$N_{cum}(>x)$ in Fig.~(1c).
The event count histogram (Fig.~1a) displays the inertial range $(x_0, x_2)$,
the minimum $(x_1)$, and maximum value $(x_2)$ of the size parameter $x$,
with a peak in the event count histogram at $x_0$ (Fig.~1a), which provides
a suitable threshold definition, because incomplete
sampling of small values $x < x_0$ is manifested by the drop of
detected events on the left side of the peak $x_0$. This definition of
a threshold $x_0$ has been proven to be very useful for characterizing
data sets with incomplete or sensitivity-limited sampling (Aschwanden 2015). 
This provides also a definition
for the inertial range $(x_0, x_2)$, which we will quantify by the
number of decades as a logarithmic ratio, i.e., $\log(x_2/x_0)$.  

An equivalent method to the cumulative size distribution is the
{\sl rank-order plot}. 
If the statistical sample is rather small, in the sense that no reasonable
binning of a histogram can be done, either because we do not have
multiple events per bin or because the number of bins is too small to
represent a distribution function, we can create a rank-order plot.
A rank-order plot is essentially an optimum adjustment to small 
statistics, by associating a single bin to every event. From an event
list of a parameter $x_i, i=1, ..., n_x$, which is generally not sorted,
we have first to generate a rank-ordered list by ordering the events
according to increasing size,
\begin{equation}
        x_1 \le x_2 \le ... \le x_j \le ... \le x_n \ , \quad j=1, ..., n \ .
\end{equation}
The bins are generally not equidistant, neither on a
linear nor logarithmic scale, defined by the difference between subsequent
values of the ordered $x_j$,
\begin{equation}
        \Delta x^{bin}_j = x^{bin}_{j+1} - x^{bin}_j \ .
\end{equation}
In a rank-ordered sequence of $n_x$ events, the probability for the largest
value is $1/n_x$, for events that are larger than the second-largest event it is
$2/n_x$, and so forth, while events larger than the smallest event occur
in this event list with a probability of unity. Thus, the cumulative
frequency distribution is simply the reversed rank order,
\begin{equation}
        N_{cum}(>x_j) = ( n_x + 1-j ) \ , \qquad j=1,...,n_x \ ,
\end{equation}
and the distribution varies from $N_{cum}(>x_1)=n_x$ for $j=1$ to
$N_{cum}(>x_n)=1$ for $j=n_x$.
We can plot a cumulative frequency distribution with $N_{cum}(>x_j)$
on the y-axis versus the size $x_j$ on the x-axis. The distribution is
normalized to the number of events $n_x$,
\begin{equation}
        \int_{x_1}^{x_n} N(x) dx = N_{cum}(>x_1)=n_x \ .
\end{equation}
We ovelay rank-order plots in each of the cumulative size distributions
in Figs.~1-6 (shown with black diamond symbols), for the $n_{ee}=10$
most extreme events in every size distribution.
If there is a {\sl ``Dragon-King''} event, it should manifest itself
by a significant offset from the best-fit cumulative size distribution
at the upper end in the zone of most extreme events near $x \lapprox x_2$,
say by a factor of two at least. The example shown in Fig.~1 exhibits
a ratio of $q_{ee}=1.03$ for the most extreme event, which is a very
small offset from the best-fit cumulative size distribution 
that does not manifest a Dragon-King event. 

\subsection{	Least-Square Fits of Size Distributions }

In order to test the consistency of the size distribution of
extreme events with the size distribution of all events we perform
least-square fits of the cumulative size distributions, such as
shown in Fig.~1c: the cumulative distribution of an observed 
parameter $x$ is shown in form of a black histogram with 
error bars $N_{cum}(>x)$, while the best-fit cumulative 
size distribution is shown with a red curve (fitted over
the range $(x_0, x_2)$ above the threshold), and the 
extrapolated range below the threshold $x_0$ is shown with
a dashed red curve. We see in Fig.~(1c) that the reduced
chi-square value is $\chi_{all}=1.46$ for all events,
while the chi-square value is $\chi_{ee}=0.56$ in the
range that contains the $n_{ee}=10$ most extreme events,
and thus is self-consistent between the two ranges, so 
that we can conclude that the extreme events do not show any 
significant deviation ($\chi \gapprox 2$) from the overall (best-fit)
thresholded power law distribution in this particular data set.

We estimate the goodness-of-fit from the uncertainties
$\sigma_{cum}(x_i)$ of the values $N_i, i=1,...,n_x$
in the fitted size distribution with the standard
reduced $\chi^2$-criterion, 
\begin{equation}
        \chi_{cum} = \sqrt{ {1 \over (n_i-n_{par})}
        \sum_{i=1}^{n_x} { [N_{cum,fit}(x_i) - N_{cum,obs}(x_i)]^2
        \over \sigma_{cum,i}^2 } }\ ,
\end{equation}
where $n_x$ is the number of bins, $n_{par}=2$ is the 
number of free parameters ($n_0, a$) of the fitted size distribution,
$N_{cum,obs}(>x_i)$ is the observed cumulative number of events
in each bin, 
$N_{cum,fit}(>x_i)$ is the fitted cumulative number of events
in each bin, 
and $\sigma_{cum}(x_i)$ is the estimated uncertainty of the
cumulative number of events,
\begin{equation}
	\sigma_{cum,i} = \sqrt{N_{cum}(x_i) - N_{cum}(x_{i+1})} \ .
\end{equation}
This definition is slightly different from the estimate in the
previous study (Aschwanden 2015), i.e., 
$\sigma_{cum,i} = \sqrt{( N_{cum,i} )}$, 
which represents a lower limit on the uncertainty only, since
the counts of events in each bin are not independent of each other
in a cumulative distribution function.
Our new approach takes only independent events into account for the
estimation of the uncertainty, which in a cumulative distribution
is the increment $[N_{cum}(>x_i) - N_{cum}(>x_{i+1})]$, 
while the remainder 
of the counts in the bins $N_{cum}(>x_{i+1}), ..., N_{cum}(>x_{n_x})$ 
do not vary in the cumulative fit of bin $N_{cum}(>x_i)$, and 
therefore do not contribute to the uncertainty $\sigma(x_i)$.

Besides the fits of the cumulative size distributions (Fig.~1c), 
we perform also fits to the differential size distributions
for consistency tests (Fig.~1b). Since the differential size 
distribution is quantified in terms of counts per bin width,
$N_{diff}(x)=n_{cts}(x)/\Delta x$, the corresponding 
$\chi^2$-criterion is, 
\begin{equation}
        \chi_{diff} = \sqrt{ {1 \over (n_x-n_{par})}
        \sum_{i=1}^{n_x} { [N_{fit,diff}(x_i) - N_{sim,diff}(x_i)]^2
        \over \sigma_{diff,i}^2 } } \ ,
\end{equation}
and the uncertainty $\sigma_{diff,i}$ in terms of Poisson statistics is,
\begin{equation}
        \sigma_{diff,i}=\sqrt{n_{cts}(x_i)} / \Delta x_i
	= \sqrt{N_{diff}(x_i) \Delta x_i)} / \Delta x_i \ .
\end{equation}
An example of such a fit is shown in Fig.~1b. Note that the size
distribution of the differential size distribution shows a straight
power law function even in the range of extreme events,
while the cumulative size distribution shows an
exponential-like drop-off towards the zero value at $x_2$,
as a consequence of the integral function defined in Eq.~(6).
We can compare now the results of the two power law fits and
find a power law slope of $\alpha_{diff,all}=1.75\pm0.02$ for
the differential size distribution (Fig.~1b), while the 
the power law slope is $\alpha_{cum,all}=1.73\pm0.02$ for
the cumulative size distribution (Fig.~1c). Now, if we
inspect the $\chi^2$-values in the range of the ten most
extreme events, we find $\chi_{cum,all}=1.46$ for all events 
(above the threshold $x_0$), and a value of $\chi_{cum,ee}=0.56$ 
for the 10 most extreme events (Fig.~1c). Therefore, not only the
two methods of differential and cumulative distributions
are self-consistent, but the  
two different ranges (of small-to-large and extreme events) 
lead to a self-consistent size distribution function also, so that
we can conclude that extreme events belong to the same
population as all other events, and that no significant deviation
from the thresholded power law distribution function is found
for this data set, which was obtained from hard X-ray peak
photon count rates in solar flares, using the {\sl Hard X-ray
Burst Spectrometer (HXRBS)} onboard the {\sl Solar 
Maximum Mission (SMM)}. The first size distributions of
these solar flare data were published in Dennis (1985)
and Crosby et al.~(1993).

Alternatively to the chosen method of least-square fitting 
of differential or cumulative size distributions, other methods 
have been used to test the consistency of an ideal power law
function with observed size distributions, such as
the maximum likelyhood estimator, the Bayesian information 
criterion, the Anderson-Darling test, or the 
Kolmogorov-Smirnov statistic (Clauset et al.~2009).
However, since no accomodation for (i) undersampling of
small events, (ii) subtration of event-unrelated background, 
or (iii) automated determination of a threshold
is provided in those alternative methods, they are 
not suitable for modeling of astrophysical data 
(Aschwanden 2015).

\subsection{	Solar Flare Observations 	}

The results of the data analysis are presented in Figs.~(2-6)
and in Table 1. While we used the generic term
{\sl ``size''} in the definition of occurrence frequency 
distributions, we should be aware that the size can be
represented by any
measure, magnitude, or physical quantity that can be
measured in a set of events, but we should not expect
to find the same power law slope for different quantities,
because there is oftern some nonlinear scaling involved.
For the solar observations analyzed here we sample
the following quantities: peak count rates of the hard X-ray
photon fluxes of flares $(P)$ (observed with the instruments 
HXRBS, the {\sl Burst and Transient Source Experiment (BATSE)}
on the {\sl Compton Gamma Ray Observatory (CGRO)}, and the
{\sl Ramaty High Energy Solar Spectroscopic Imager (RHESSI)}
spacecraft), total counts (or fluences) 
of hard X-ray fluxes $(C)$,
flare durations $(T)$, emission measure $(EM)$ in soft X-rays and EUV,
thermal flare energies $(E_{th})$, 
nonthermal flare energies $(E_{nth})$, 
dissipated magnetic energies 
$(E_{mag})$, flare volumes $(V)$, flare areas $(A)$, and flare 
length scales ($L$). The energetic parameters have been calculated 
in recent statistical studies on the global energetics of solar flares 
(Aschwanden et al.~2014, 2015, 2016).

The results of forward-fitting differential and cumulative 
size distributions of different events 
are summarized in Table 1, which includes:
background counts (BG), the number of all events (N),
the number of events above threshold (NT),
the logarithm of the inertial range $\log(x_2/x_1)$ (IR),
the threshold value ($x_0$),
the power law slope from fitting differential ($\alpha_{diff}$)
and cumulative occurrence frequency distributions ($\alpha_{cum}$), 
the $\chi^2$-values of differential ($\chi_{diff}$) and
cumulative size distributions ($\chi_{cum}$),
and the mean $\chi^2$-values in the range of the
10 most extreme events ($\chi_{ee}$).

We determined the background level by minimizing the 
goodness-of-fit $\chi^2$-criterion as a function of the
background level (see Fig.~7 in Aschwanden et al.~2015).
It turned out that only the HXRBS data needed background
subtraction (by 41 and 58 cts s$^{-1}$), while
the BATSE and RHESSI flare catalogs provide 
background-subtracted flux values.

The total number of events ($N$ in Table 1) encompasses all
sampled events, but only those (NT) above the threshold $x_0$  
could be used in the fitting of size distributions. The inertial
range is characterized by the logarithmic ratio
$\log(x_2/x_0)$, which ranges from 2 to 5 decades.
Those with the largest inertial ranges and largest number of
events provide the most accurate fits of size distributions.

\subsection{	Self-Consistency of Size Distributions		}

A first important test is the self-consistency between
the two different methods, i.e., the fitting of the
differential and cumulative size distributions. We can
simply compare the resulting power law slopes, which are
labeled as $\alpha_{diff}$ and $\alpha_{cum}$ in Table 1.
We find a satisfactory agreement between the two methods
within the derived uncertainties of a few percents,
as it can be seen for all data sets with large statistics
(Table 1: $N \approx 10^3-10^4$ events). This corroborates the
self-consistency between the two methods. The high accuracy
is mostly achieved by modeling the size distribution with
a thresholded power law distribution function (Eqs.~3 and 6), 
which is more suitable to represent the data than an ideal 
power law function.

A second important test is the comparison of the 
goodness-of-fit between the differential and cumulative
size distributions, labeled as $\chi_{diff}$ and $\chi_{cum}$
in Table 1. Among the 9 fitted size distributions of 3 solar
observables (P, C, T) with 3 different instruments each
(HXRBS, BATSE, RHESSI), shown in Fig.~2, we find that only one data set
is consistent with the fitted (differential and cumulative)
size distributions, namely P of HXRBS, 
with $\chi_{diff} \approx 2$ and
$\chi_{cum} \lapprox 2$, while the other 8 data sets exhibit
significant mismatches $\chi_{diff} \approx \chi_{cum}
\approx 3-13$. This indicates that the large number
statistics of these data sets is sensitive to apparently 
little but significant deviations from power laws. 
These deviations can clearly be recognized in the poor 
values $\chi_{diff,all}$ and $\chi_{cum,all}$ in the
plots of Figs.~(2b-2i). There is a
tendency that peak count (P) size distributions match a power
low function closest, while the total counts (C) and the 
flare durations (T) deviate significantly. The reason for the latter
property could be due to a violation of the principle of
time scale separation that is required in SOC models
(Aschwanden et al.~2016), caused by confusion between
overlapping short and long-duration flares.

\subsection{	Energetic Solar Flare Parameters	}

The flare observables ($P, C, T$) are easiest to measure 
and have often been used in modeling of SOC models. However,
since these observables may not be a good proxy for the 
representation of the energy contained in SOC avalanches,
we inspect also the size distribution of spatial and 
flare parameters, such as ($L, A, V, E_{mag}, E_{th}, E_{nth}, EM$),
which have been determined under the scope of global
energetics in solar flares in a series of ongoing studies
(Aschwanden et al.~2014, 2015, 2016). The cumulative
size distributions are presented in Figs.~3 and 4.
Since the number of events, for which energetic parameters
have been determined, includes M and X GOES-class flares
only, we deal with small-number statistics  
in the order of $N\approx 100-300$ events per data set.
Consequently, the resulting fits yield acceptable 
values for the goodness-of-fit, i.e., $\chi^2 \lapprox 2$,
but these small data sets have insufficient sensitivity 
to measure deviations from power law functions. 
Nevertheless, since they contain large flares only, they
are suitable to detect the outliers of extreme events.

The only obvious deviation from a power law is detected 
for the size distribution of nonthermal energies (Fig.~4b),
which is suspected to arise due to an instrumental
irregularity, such as pulse pile-up in RHESSI data.

\subsection{		Extreme Events			}

We investigate now the so-called Dragon-King hypothesis
(Sornette 2009; Sornette and Ouillon 2012). This hypothesis 
suggests that the most extreme events may belong to a
different population of event size distributions than
smaller or even large events. In this study we define
three different parameters to characterize Dragon-King
events: (i) A range of largest extreme events in a 
data set, which we set arbitrarily to $n_{ee}=10$ here;
(ii) The ratio of the observed size $x_{max,obs}=x_2$ to
the expected size $x_{max,fit}$ of the most extreme
event, based on the cumulative size distribution fit,
i.e., $q_{ee}=x_{max,obs}/x_{max,fit}$, 
(which we may call the {\sl extreme event excess factor}),
and (iii) the goodness-of-fit $\chi_{ee}$ of the cumulative
size distribution fit in the range of the largest
$n_{ee}=10$ events. This three values are given in
the lower right corner of each panel in Figs.~1-6 and Table 1.

As an example we consider Fig.~1, where the goodness-of-fit
in the extreme event part of the size distribution, 
i.e., $\chi_{ee}=0.56$, which compares well with the
value obtained from fitting all events above the threshold,
i.e., $\chi_{all}=1.46$, which does not yield any
evidence for a different population. A second test
is the extreme event excess factor, which for this event
is $q_{ee}=1.03$ and yields no evidence for a Dragon-King
event either (Fig.~1c).  

We list all evaluated extreme event excess factors 
$q_{ee}$ in Table 1 (also given in the panels of Figs.~1-6).
Among the solar flare parameters, all those factors
vary in the range of $q_{ee}\approx 0.9-1.9$, with a
single outlier of $q_{ee}=3.08$ (for one flare volume
data set, see Fig.~3c), out of 23 solar data sets.
Therefore, we see no indication that solar extreme events 
belong to different populations of event distributions. 

The estimation of an unknown tail distribution function
can be derived with tools from {\sl extreme value theory},
such as with the {\sl Pickands-Balkema-de Haan theorem}
(Balkema and de Haan 1974; Pickands 1975), 
the {\sl Fisher-Tippett-Gnedenko theorem}
(Fisher 1930), or as described in Gumbel (1958),
which converge to three possible distributions 
(the Gumbel distribution, the Fr\'echet distribution, or
the Weibull distribution). In our case, however, the
cumulative form of the {\sl thresholded power law size
distribution} (generalized Pareto) function is 
our preferred model of choice, due to its ability of taking
incomplete sampling, background subtraction, and finite
system size into account, and thus is known a priori, 
so that it does not need to be estimated with 
extreme value theory, as it would be needed in the case
of an unknown tail distribution function.
The theoretically defined
distribution is then fitted to the observed 
distributions directly, and the {\sl extreme event 
excess factor} $q_{ee}$ is conveniently quantified in
terms of standard deviations in common $\chi^2$-statistics. 

\subsection{	Stellar Flare Observations 		}

Impulsive flaring with rapid increases in the brightness
in UV or EUV has been observed for a number of so-called
flare stars, such as AD Leo, AB Dor, YZ Cmi, EK Dra, or
$\epsilon$ Eri. These types of stars include cool M dwarfs,
brown dwarfs, A-type stars, giants, and binaries in the
Hertzsprung-Russell diagram. Most of these stars are
believed to have hot soft X-ray emitting coronae, similar
to our Sun (a G5 star), and thus magnetic reconnection
processes are believed to operate in a similar way
as on our Sun (see, e.g., review by G\"udel 2004).

We complement the solar flare data sets with three data sets of
stellar flares, two recorded with KEPLER and one observed
with the EUVE telescope, with the cumulative size 
distributions shown in Figs.~(4g, 4h) and Fig.~(5). 
The fit of a cumulative power law distribution 
shown in Fig.~(4g) produces
a size distribution with a power law slope of $\alpha
=1.68\pm0.12$ for the peak counts, which is similar to 
the slope observed in solar flares, i.e., 
$\alpha=1.70\pm 0.02$ with HXRBS (Fig.~2a),
$\alpha=1.81\pm 0.02$ with BATSE (Fig.~2b), and
$\alpha=1.86\pm 0.02$ with RHESSI (Fig.~2c). 
The goodness-of-fit for extreme events is $\chi_{ee}=1.45$,
which is similar to $\chi_{all}=1.26$ for all events (Fig.~4g).
 
A second data set with stellar flare data from KEPLER,
derived from the bolometric energy, is shown in Fig.~(4h),
yielding a power law slope of $\alpha_{bol}=2.55\pm 0.06$,
which is similar to the value expected for flare areas 
$\alpha_A=2.33$ (Table 2). Since the largest contributions
to the bolometric intensity come from the solar or stellar
photospheric surface area, it makes sense the its size 
distribution scales with the flare area (rather than
with a 3-D volume).

In Fig.~5 we present cumulative size distribution fits of 
12 cool (type F to M) stars, observed by Audard et al.~(2000),
with typically 5-15 flare events per star.
For most of these stars, the flare intensities follow closely
the fitted cumulative distribution all the way to the most
extreme event, except for two cases (out of 12):
HD 2726 (F2 V) (top left panel) and CN Leo, 1994 (M6 V)
(bottom middle panel), which could qualify as outliers
(or Dragon-King events). The power law slopes inferred
in this sample vary strongly within a range of 
$\alpha \approx 2.0\pm0.4$, which tend to be somewhat
higher than the mean values of solar flare size distributions,
but are not reliable in such small number statistics.

In summary, stellar flares fit cumulative
size distributions at the upper cutoff (within an energy
range of $\approx 0.8-1.4$ decades), but the
power law slopes appear to be slightly steeper than
observed in solar flares, and outliers of extreme events 
occur only rarely (in 2 out of 12 cases).

\section{	DISCUSSION 				}

\subsection{	Astrophysical Predictions of Power Law Slopes }

While we are interested in the investigation of deviations 
from ideal power laws, which is a second-order effect, we should
consider first the physical justification of the existence
of power law size distributions in the first place.  
The original discoverers of {\sl self-organized criticality
(SOC)} models noted power law-like size distributions with power law
slopes of $\alpha_x \approx 2$ of the sizes $x$ of SOC avalanches, 
as well as power law slopes of $\alpha_T \approx 2$ for the durations 
$T$ of SOC avalanches, obtained in cellular automaton simulations
(e.g., Pruessner 2012).
A more physical
approach of modeling and predicting power law slopes of
size distributions observed from astrophysical SOC systems
has been put forward in terms of the so-called 
{\sl Fractal-Diffusive (FD-SOC)} model (Aschwanden 2013, 2014,
2016). We juxtapose theoretical predictions of this model
and mean observational values in Table 2.

The most fundamental assumption in the FD-SOC model is the
{\sl scale-free probability conjecture}, which directly predicts 
a power law size distribution function of geometric length 
scales $L$ from first principles, i.e., 
\begin{equation}
	N(L) dL = L^{-S} \ dL \ ,
\end{equation} 
where $S$ is defined as the Euclidean dimension of the SOC system
(with possible values of $S=1,2,3$).
From this fundamental power law distribution, other power
law size distributions of any physical size parameter $x$ 
that has a linear $(x \propto L)$, or a nonlinear relationship 
$x(L) \propto L^p$ to the length scale $L$ (with power index $p$),
can be derived. The following scaling laws have been used
in the calculation of the power law indices listed in Table 2:
\begin{eqnarray}
 		A & \propto & L^{D_2} \\
		V & \propto & L^{D_3} \\
		L & \propto & T^{\beta/2} \\
		F & \propto & V^\gamma \propto L^{D_S \gamma} \\
		P & = & F_{max} \propto V_{max}^\gamma \propto L^{S \gamma} \\
		E & = & F \ T  \propto L^{(\gamma D_S + 2/\beta)} 
\end{eqnarray}
where $D_2$ is the area ($A$) fractal dimension,
$D_3$ is the volume ($V$) fractal dimension,
$\beta$ is the spatio-temporal diffusion coefficient
for the time duration ($T$) of a SOC avalanche 
with classical diffusion $(\beta=1)$,
$\gamma$ is the flux-volume scaling of observed fluxes $F$ of
astrophysical sources,
$V_{max}$ is the maximum fractal volume, which obeys the Euclidean
dimension $S$ of the SOC system, 
$P=F_{max}$ is the peak or maximum flux, 
and $E=F\ T$ is the fluence, defined
as the product of the flux $F$ and time duration $T$.

The mean fractal dimension is estimated from the average of the
minimum ($D_{S,min} \approx 1$) and maximum $D_{S,max}=S$ values, 
\begin{equation}
	D_S \approx {(D_{S,min}+D_{S,max}) \over 2} 
	= {(1+S) \over 2} \ , 
\end{equation}
which yields $D_2=(1+2)/2=3/2$ and $D_3=(1+3)/2=2$ for SOC
models that occupy either an Euclidean 2-D or 3-D dimensional 
space. 

Since all relationships given in Eq.~(15-21) are expressed in terms
of the variable $L$, the power law indices $\alpha_x$ for the
size distribution of the parameters $x=A, V, T, F, P, E$
can be derived by substitution of variables,
i.e., $N(x) dx =\ N[x(L)]\ |dx/dL]\ dL \propto x^{-\alpha_x}$,
yielding the power law indixes $\alpha_x$,  
\begin{eqnarray}
 		 \alpha_L &=& S \approx 3 \\
 		 \alpha_A &=& 1 + (S - 1)/D_2 \approx 7/3 \\
 		 \alpha_V &=& 1 + (S - 1)/D_3 \approx 2 \\
 		 \alpha_T &=& 1 + (S - 1) \beta/2 \approx 2 \\
 		 \alpha_F &=& 1 + (S - 1)/(\gamma\ D_S) \approx 2 \\
 		 \alpha_P &=& 1 + (S - 1)/(\gamma\ S) \approx 5/3 \\
 		 \alpha_E &=& 1 + (S - 1)/(\gamma\ D_S + 2/\beta ) \approx 3/2 \ ,
\end{eqnarray}
where the approximative numerical values are obtained by inserting
$S=3, \beta=1, \gamma=1, D_2=1.5, D_3=2$. 

In Table 2 we juxtapose these theoretical power law slope values
with the observed values. We see that there is mostly a good agreement
(within a few percents) for solar flare size distributions with
large statistics ($N \approx 10^3-10^4$), such as for peak fluxes
($\alpha_P$), total counts ($\alpha_C$), and flare durations ($\alpha_T$),
while others with small number statistics show larger deviations from
the predictions, such as for length scales ($\alpha_L$), areas ($\alpha_A$), 
and volumes ($\alpha_V$). 

New results are obtained here from the power law slopes of energetic flare
parameters. The dissipated magnetic energies $E_{mag}$ are found to
have power law indices that are close to the values of flare areas,
i.e., $\alpha_{mag} = 2.8\pm 0.2$ versus $\alpha_A \approx 2.3$, which could be
explained by the fact that the dissipation of free magnetic energies
is concentrated in the chromospheric flare ribbons, which have a
horizontal area-like surface geometry. This is also true for
the bolometric flare energies observed in stellar flares, 
i.e., $\alpha_{bol} = 2.6 \pm 1.0$ versus $\alpha_A \approx 2.3$, 
since the bolometric brightness is mostly irradiated in 
the photosphere of stellar surfaces. Also for the thermal emission
measure we find similarity with area-like features,
i.e., $\alpha_{EM} = 2.8 \pm 0.2$ versus $\alpha_A \approx 2.3$,
which suggests that the soft X-ray emission measure in solar flares
has an area-like extension with relatively low altitudes. 
For thermal flare energies, for which we expect mostly a volume
dependence,
\begin{equation}
	E_{th} = 2 n_e k_B T_e V \propto V \propto L^{D_3} \ ,
\end{equation}
since the electron density and the electron temperature vary
much less than the flare volume. Indeed we find values of
$\alpha_{E_{th}} = 2.2 \pm 0.1$ that are similar to the volume
power law index, $\alpha_V \approx 2$. For nonthermal flare
energies we find a close coincidence of the power law slope
of $\alpha_{E,nth}=1.57 \pm 0.11$, which is close to the
power law slopes of fluences, $\alpha_{C} \approx 1.50$,
which is expected because both the fluences
($E=F\ T$) and the nonthermal energies, $E_{nth} 
= \int e_{nth}(t)\ dt$, are time-integrated flare quantities.

These comparisons of power law indices thus reveal interesting
relationships that are relevant for determining the physical 
scaling laws that govern solar flares.

\subsection{	Outliers or Dragon-King Events		}

From the 23 solar (Figs.~3, 4) and 14 stellar data sets 
(Figs.~4, 5) we found little evidence for outliers among
the most extreme events in each data set. The largest outliers
for the most extreme events are found for a small data set
($N=172$) of flare volumes with an excessive extreme event factor of
$q_{ee}=2.75$ (Fig.~3c), for a small stellar flare data set  
($N=209$) with $q_{ee}=3.55$ (Figa.~4g), and for the two stars
HD 2726 and CN Leo with even less statistics
($N \approx 15$) (Fig.~5). The fact that all
these most extreme events occur in small data sets, while 
larger data sets with $N \gapprox 10^3-10^4$ events (Fig.~2)
reveal no excesses larger than $q_{ee} \lapprox 2.0$, 
provides little evidence for the existence of 
super-extreme events that are not part of the ``normal''
population of small and large flare events. 
Hence, we conclude that extreme events of solar and stellar
flares are largely scale-free (over an inertial range of
$\approx 3-4$ decades) in their volumes, fluxes, or energies.

What does it mean that we find no significant outliers in the
cumulative size distributions of solar and stellar flares?
As the size distribution of length scales reveals in Fig.~(3a),
solar flares have length scales of $L \approx 10-200$ Mm,
which at the upper limit matches the maximum size of active
regions on the solar surface. The largest possible flare that
extends over the entire solar surface would have a 
circumference of $2 \pi \times 696$ Mm, which is apparently
impossible since we never observed a flare with a size larger
than $L \lapprox 200$ Mm. One effect that could help to increase
the maximum flare size is the case of {\sl sympathetic flaring},
which involves a magnetic coupling between adjacent (or even
remote) active regions. However, the larger an unstable 
magnetic field configuration is, the smaller is the filling
factor of magnetic energy and heated plasma with respect
to the size of a flaring region. Consequently, the envisioned
process of amplification and synchronization promoted in the
Dragon-King hypothesis (Sornette 2009; Sornette and Ouillon 
2012) seems to be questionable for solar and stellar flares. 

\subsection{	Deviations from Power Laws		}

It has been argued that real data deviate from ideal power law
size distributions, as they have been reproduced by cellular
automaton simulations (e.g., Pruessner 2012). Some sceptics
went even as far as to deny the existence of power laws at all
(Stumpf and Porter 2012). Traditional least-square fitting
techniques have been criticized to be inaccurate, while
maximum-likelyhood fitting methods with goodness-of-fit tests
based on the Kolmogorov-Smirnov statistic were preferred
(Clauset et al.~2009). However, we have demonstrated here
and earlier (Aschwanden 2015) that
least-square fitting methods of both differential and
cumulative size distributions yield self-consistent fits
of the power law slope, as long as data truncation, 
undersampling of data in small events below some threshold, 
subtraction of event-unrelated background
noise, and the steep fall-off at the upper bound of the cumulative
size distribution is properly modeled. 
In other words, a straight power law function 
represents an over-simplified model for most cases, while 
inclusion of the effects enumerated above is essential in 
defining adequate models for fitting statistical distributions. 

Now, with the availability of large data sets (with $N \approx
10^5-10^8$ events), we are becoming much more sensitive to the
smallest deviations from ideal power laws. So it is no
surprise that deviations from power laws become the rule 
rather than the exception. The deeper reason for these 
power law deviations is the inhomogeneity of the medium
and the physical conditions. It is instructive to consider
the sandpile analogy of SOC systems. Although the cone shape
of a sandpile appears to have a constant slope or angle of
repose, when viewed from distance, closer inspection reveals 
macroscopic channels, erosions, bumps, and dents, on top of the
microscopic fine structure. As an illustration we mention 
some empirical data sets (chosen by Clauset et al.~2009) that
are shown in Fig.~6 and described in Appendix A. Some data sets
clearly reveal a complete inadequacy to fit any power law 
function to data with a broken power law 
(Fig.~6b: wildfires; Fig.~6c: cities).
The case of weblinks has such a tremendous statistic
($N=1.35 \times 10^8$ events) that it is highly sensitive to
the smallest deformations from an ideal power law, i.e., 
amounting to 
$\chi_{all}=223$ for a slight deficit near $x \approx 10^5-10^6$
(Fig.~6h). A Dragon-King event with an excess of $q_{ee} \approx 5$ is
indicated in the data set of terrorism (Fig.~6e).

Significant deviations from ideal power law size distributions
have also been detected in stellar flares, e.g. in the Kepler-based
study of Sheikh et al.~(2016), and most of the analyzed stars
were found to exhibit no evidence for SOC avalanching. It is argued
that these stellar flares are subject to ``tuned criticality''
rather than ``self-organized criticality'' (Sheikh et al.~2016). 

\subsection{	Physical Mechanisms of SOC Avalanches	}

The observations of SOC avalanches exhibit power law-like
size distributions, which constrain (scale-free) scaling laws of 
physical parameters (see Section 3.1), but do not directly reveal 
the physical mechanism of SOC processes as such. The power law slopes
of the size distributions of solar and flare data have been
found to be consistent with the scale-free probability conjecture
(Eq.~15), fractal-diffusive transport (Eq.~18), and with
EUV and soft X-ray emission (flux or intensity) that is 
quasi-proportional to the fractal volume of an avalanche event
(Eq.~20). These scaling laws are consistent with standard
solar flare models, which are driven by a magnetic reconnection 
process in the solar corona (Lu and Hamilton 1991). 
However, the observed scaling laws are also consistent with
other physical mechanisms, such as turbulent photospheric
convection, which may be (stochastically) coupled with coronal
reconnection events (Uritsky et al.~2013; Uritsky and Davila 2012;
Knizhnik et al.~2017). Additional interpretations of physical
mechanisms operating in SOC systems can be found in
Aschwanden et al.~(2011, 2016).  

The interpretation of SOC systems in terms of physical 
mechanisms is beyond the scope of this study, but the fact
that we find virtually no outliers or Dragon-King events
implies that only one single physical mechanism with
appropriate properties is needed to explain the observed 
power law size distributions. A dual energy dissipation system
is likely to produce broken power law distributions.
However, the same physical process, such as a magnetic 
reconnection process, can have multiple secondary energy
dissipation processes, including nonthermal particle acceleration
(hard X-ray fluxes), thermal flare plasma heating (soft X-ray
fluxes), launching of a coronal mass ejection (white light
emission), or electron beam formation (radio emission).
Quantitative energy budget calculations have revealed 
energy closure for the primary flare-dissipated magnetic energy
and secondary energy dissipation mechanisms (Aschwanden
et al.~2017). All energy that is dissipated during a solar
flare is supplied by the free (magnetic) energy in an active
region. Interestingly, multi-scale intermittent dissipation
with power law-like SOC properties has also been detected in 
Quiet Sun regions at spatial scales of $\gapprox 3$ Mm, 
controlled by turbulent photospheric convection (Uritsky
et al.~2013). It appears that we observe a coexistence 
of coronal SOC systems (in active regions) and photospheric
SOC systems (in Quiet Sun regions).

\section{	CONCLUSIONS 				}

In this study we search for outlier events in astrophysical
data sets from solar and stellar flares. Such an endeavour
is motivated by the {\sl ``Dragon-King hypothesis''} (Sornette
2009; Sornette and Ouillon 2012), which suggests that the most
extreme events in a statistical distribution may belong to a
different population, and thus may be generated by a different
physical mechanism, than the ubiquitous power law distributions 
commonly observed in nonlinear systems driven by self-organized
criticality. For this project
we extracted 23 data sets of solar flare data (from HXRBS/SMM,
BATSE/CGRO, RHESSI, AIA/SDO, HMI/SDO), 12 data sets from
stellar flares (from EUVE and KEPLER), and another 8 data sets
of non-astrophysical data. We define a thresholded power law
distribution, which includes the effects of data truncation, 
undersampling of small events below a threshold, subtraction 
of event-unrelated background noise, and the steep fall-off 
at the upper bound of cumulative size distributions. We 
tested the accuracy of our size distribution forward-fitting
method for both the differential and the cumulative occurrence
frequency size distributions, as well as using the 
method of rank-order
plots. Our least-square fitting techniques focus mostly on 
detecting deviations from ideal power laws (by means of the
$\chi^2$-criterion) and on detecting outliers of the most
extreme events in each statistical distribution.
In the following we summarize the major conclusions.

\begin{enumerate}

\item{{\sl Identifying Dragon-King events is important for 
accurate predictions and forecasting of the largest catastrophes.}  	
For solar and stellar flare data, the frequency and size
of the most extreme events have generally been established
from extrapolating power law distributions, which may
under-estimate the largest events, if such outlier events
belong to a different population and are generated by different
physical mechanisms. Prediction and forecasting of large events 
generated by solar flares or coronal mass ejections have been
recognized to play a central role in {\sl space weather 
predictions} (e.g., Schrijver 2007, 2009; Gallagher et al. 2002;
Georgoulis and Rust 2007; Leka and Barnes 2007; Barnes et al.~2007;
Barnes and Leka 2008; Bloomfield et al.~2002; Aschwanden 2019,
Section 16.8 and 16.9 and references therein).}

\item{{\sl Dragon-King events are theoretically possible in solar 
	and stellar flares}, e.g., in the case of sympathetic flaring. 
	Sympathetic flares are a pair of flares that occur almost
	simultaneously in different active regions, not by
	chance, but because of some physical connection (for 
	instance by propagating Alfv\'enic waves). 
	Statistical evidence for
	sympathetic flaring has been established for 48 events,
	some of them consisting of trans-equatorial loops
	(Moon et al.~2002). Another example is a quadrupolar 
	configuration with two magnetic flux ropes that are
	located within a pseudo-streamer, which have
	been observed to lead to two consecutive reconnections 
	and eruptions, a scenario for twin-filament eruptions
	that can explain coupled sympathetic eruptions also 
	(T\"or\"ok et al.~2011). Coupling flares in adjacent
	regions could double up the joint flare volumes and
	energies, if both active regions have similar sizes.}

\item{{\sl Dragon-King events are not common in 
	solar and stellar flares.} In our analysis we measured
	excessive extreme event factors $q_{ee}$ for each
	data set, which characterizes the offset of outliers
	by the ratio of the observed most extreme event
	to the expected most extreme size based on a
	cumulative size distribution fit. We found that this
	factor amounts generally to less than a factor of
	$q_{ee} \lapprox 2$, in 21 out of the 25
	investigated solar and stellar size distributions.
	The four cases possibly containing Dragon-King
	events were found for solar flare volumes ($q_{ee}=3.08$),
	stellar flare peak counts ($q_{ee}=3.55$), and in 
	flares from the red dwarf stars HD 2726 and AD Leo (Fig.~5).
	However, Dragon-King events have been found in
	other, non-astronomical data sets (Fig.~6), such as in 
	earthquakes ($q_{ee}=2.82$), power blackouts
	($q_{ee}=3.35$), and terrorism ($q_{ee}=6.15$).}

\item{{\sl Extreme events of solar and stellar flares are scale-free.}
	The fact that almost all extreme events of solar and stellar flares
	fit a power law size distribution with a small value of the
	excessive extreme event factor ($q_{ee} \lapprox 1.5$)
	indicates that solar and stellar flare processes are
	scale-free (within the inertial range), and thus can be
	modeled with the same power law (SOC) model for small or large 
	events. This implies also that small and large flare events
	are subject to the same physical mechanism, most likely 
	caused by a magnetic reconnection process in active regions, 
	driven by shearing and twisting of the magnetic field.}	

\item{{\sl Deviations from power laws are common in large-number 
	statistical distributions.} Adversely, distributions with
	acceptable power law fits are more likely to be found in 
	size distributions 
	with small-number statistics. The accuracy of fitting
	power law slopes scales with the square root of the number
	of events, and hence the measurement of deviations from
	ideal power laws is a matter of instrumental sensitivity.
	However, large-number statistics ($N \gapprox 10^4$) requires
	comprehensive flare catalogs or automated pattern recognition
	codes, which is the reason why we have a high accuracy of
	power law fits for extensive data sets only, such as for 
	peak counts ($P$), total counts ($C$), or time durations ($T$),
	while energetic parameters could only be measured in large
	(GOES M and X-class) flares, yielding small-number statistics
	($N \gapprox 10^2$) with a lower degree of accuracy.}

\item{{\sl Accurate size distributions provide tests of physical 
	scaling laws in solar and stellar flares.} 
	The fractal-diffusive SOC model predicts a set of
	power law slopes (for avalanches in a SOC system) that
	can straightforwardly be compared with the corresponding
	observed size distributions. Consequently, the obtained
	power law slopes provide consistency tests of the
	underlying physical scaling laws. In this study we
	discovered new correlations that indicate that the 
	dissipated magnetic energies 
	($\alpha_{mag} \propto \alpha_A$), flare emission measures
	($\alpha_{EM} \propto \alpha_A$), and bolometric energies
	($\alpha_{bol} \propto \alpha_A$) scale with the flare area,
	that the thermal flare energy scales with the flare volume
	($\alpha_{th} \propto \alpha_V$), and that the nonthermal
	energy scales with the total (time-integrated) hard X-ray
	counts ($\alpha_{nth} \propto \alpha_C$). The reason why
	these flare scaling laws can be inferred mostly from the
	flare volume $V$ or area $A$ is the high nonlinearity of these
	geometric parameters ($V \propto L^3$, $A \propto L^2$),
	while other physical parameters (such as the electron
	density $n_e$ or temperature $T_e$) scale only linearly
	in the definitions of the thermal energy and pressure.}
\end{enumerate}

There are a lot of new questions that arise from this study that 
can be addressed in follow-on work: What physical conditions
are different in Dragon-King events? Does sympathic flaring
occur in the most extreme events? Which stars produce outliers
and what is different in the flare productivity of these stars?
If we increase the size of the data sets to large-number 
statistics, do we obtain more accurate power law slopes that
converge from different instruments? If we improve the statistics,
do the power law indices in physical scaling laws converge to the
theoretically predicted values? Can stellar power laws be used
to tell us which flare phenomena are controlled by the flare
size and their relationship to the depth of the stellar 
convection zone?

\subsection*{	APPENDIX A: Empirical Data 		}

In the spirit of interdisciplinary research, let us visit 
eight non-astrophysical data sets of 
empirical data (Fig.~6, Table 1), which have been compiled and
scrutinized for deviations from ideal power laws by 
Clauset et al.~(2009). If we investigate the goodness-of-fit
$\chi^2$-values of all events in each data set, we find 
significant deviations from ideal power laws for most of the cases, 
except for blackouts (Fig.~6e: $\chi_{cum}=0.83$), but this phenomenon 
has also the smallest statistics ($N=210$), and thus is not
sensitive to small deviations from power laws.
On the other hand, if we consider just the average goodness-of-fit
$\chi^2$-values among the 10 most extreme events, we find consistency with
thresholded power law distributions for 
earthquakes (Fig.~6a: $\chi_{ee}=1.45$),
blackouts (Fig.~6d: $\chi_{ee}=0.63$),
terrorism (Fig.~6d: $\chi_{ee}=0.43$).
surnames (Fig.~6g: $\chi_{ee}=1.34$), and
weblinks (Fig.~6h: $\chi_{ee}=1.07$).
These results differ somewhat from those obtained in Clauset et al.~(2009),
although they also find most of the data are not consistent with
an ideal power law distribution, 
which is expected for fitting different functions to the observed 
distributions, such as the thresholded Pareto-type power law 
function used here. The evaluation
of power law deviations in Clauset et al.~(2009) is subject to
an arbitrary truncation of small events, while our method 
derives a threshold automatically, which is important, because 
the choice of (scale-free) inertial range boundaries affects 
the outcome of the power law slope value most.

\bigskip
{\sl Acknowledgements:} 
This work was stimulated by the organizers of a workshop on
{\sl ``Mechanisms for extreme event generation'' (MEEG)} at the
Lorentz Center at Snellius, Leiden, The Netherlands, July 8-12, 2019,
organized by Drs. Norma Bock Crosby, Bertrand Groslambert,
Alexander Milovanov, Jens Juul Rasmussen, and Didier Sornette. 
The author acknowledges the hospitality and partial support of two
previous
workshops on ``Self-Organized Criticality and Turbulence'' at the
{\sl International Space Science Institute (ISSI)} at Bern, Switzerland,
during October 15-19, 2012, and September 16-20, 2013, as well as
constructive and stimulating discussions with
Sandra Chapman,
Paul Charbonneau,
Aaron Clauset,
Norma Crosby, 
Michaila Dimitropoulou,
Manolis Georgoulis,
Stefan Hergarten,
Henrik Jeldtoft Jensen,
James McAteer,
Shin Mineshige,
Laura Morales,
Mark Newman,
Naoto Nishizuka,
Gunnar Pruessner,
John Rundle,
Surja Sharma,
Antoine Strugarek,
Vadim Uritsky,
and Nick Watkins.
This work was partially supported by NASA contracts NNX11A099G
``Self-organized criticality in solar physics'', 
NNG04EA00C of the SDO/AIA instrument, 
and NNG09FA40C of the IRIS instrument.
\clearpage

\section*{References} 
\def\ref#1{\par\noindent\hangindent1cm {#1}}

\ref{Aschwanden, M.J.: 2004, {\sl Physics of the Solar Corona. 
	An Introduction}, Berlin: Springer and Praxis, p.216.}
\ref{Aschwanden, M.J. 2011,  {\sl Self-Organized Criticality in 
	Astrophysics. The Statistics of Nonlinear Processes in 
	the Universe}, Springer-Praxis: New York, 416p.}
\ref{Aschwanden, M.J., Xu, Y., and Jing, J. 2014, ApJ 797:50.}
\ref{Aschwanden, M.J., Boerner, P., Ryan, D., Caspi, A.,
	McTiernan, J.M., and Warren, H.P. 2015, ApJ 802:53.}
\ref{Aschwanden, M.J. 2015, ApJ 814:19.}
\ref{Aschwanden, M.J. (ed.) 2013, {\sl Self-Organized Criticality 
	Systems}, Open Academic Press, Berlin, Warsaw, 483p.}
\ref{Aschwanden, M.J., Crosby, N., Dimitropoulou, M., Georgoulis, M.K., 
	Hergarten, S., McAteer, J., Milovanov, A., Mineshige, S., 
	Morales, L., Nishizuka, N., Pruessner, G., Sanchez, R., 
	Sharma, S., Strugarek, A., and Uritsky, V. 2016,
 	SSRv 198, 47.}
\ref{Aschwanden, M.J., Caspi, A., Cohen, C.M.S., Holmen, G.,
	Jing, J., Kretzschmar, M., Konter, E.P., McTiernan, J.M.
	2017, ApJ 836, 17.}
\ref{Aschwanden, M.J. 2019, {\sl New Millennium Solar Physics},
 	Astrophysics and Space Science Library, Volume 458, 
	ISBN 978-3-030-13954-4, Springer Nature Switzerland AG; 
	DOI: 10.1007/978-3-030-13956-8, (Section 11.8).}
\ref{Audard M., G\"udel, M., Drake, J.J., and Kashyap V.I. 2000,
	ApJ 541, 396.}
\ref{Bak, P. 1996, {\sl How nature works : the science of 
	self-organized criticality}, Copernicus, New York.}
\ref{Balkema, A. and de Haan L. 1974, Annals of Statistics 2, 792.}
\ref{Balona, L.A. 2015, MNRAS 447, 2714.}
\ref{Barnes, G., Leka, K.D., Schumer, E.A., et al. 2007,
        Space Weather 5/9, S09002.}
\ref{Barnes, G. and Leka, K.D. 2008, ApJ 688, L107.}
\ref{Baro, J., Corral, A., Illa, X., Planes, A.,, Salje, E.K.H.,
	Schranz, W., Soto-Parra, D., and Vives, E. 2013,
	Phys. Rev. Lett. 110, 088702.}
\ref{Bloomfield, D.S., Higgins, P.A., McAteer, R.T.J., et al. 2012,
        ApJ 747, L41.}
\ref{Charbonneau, P., McIntosh, S.W., Liu, H.L., and Bogdan,T.J.
 	2001, SoPh 203, 321.}
\ref{Clauset, A., Shalizi, C.R., and Newman, M.E.J. 2009,
	SIAM Rev. 51/4, 661.}
\ref{Crosby, N.B., Aschwanden, M.J., and Dennis, B.R. 1993,
	SoPh 143, 275.}
\ref{Dennis, B.R. 1985, SoPh 100, 465.}
\ref{Fisher, R.A. 1930, {\sl The Genetical Theory of Natural Selection},
	Oxford University Press, Oxford.}
\ref{Gallagher, P.T., Moon, Y.J., and Wang, H. 2002, SoPh 209, 171.}
\ref{Georgoulis, M.K. and Rust, D.M. 2007, ApJ 661, L109.}
\ref{G\"udel, M. 2004, Astron. Astrophys. Rev. 12(2-3), 71.} 
\ref{Gumbel, E.J. 1958, {\sl Statistics of extremes}, Columbia
	University Press, Hew York.}
\ref{Hergarten, S. 2002, {\sl Self-Organized Criticality in 
	Earth Systems}, Springer, Berlin.}
\ref{Hosking, J.R.M. and Wallis, J.R. 1987, Technometrics 29, 339.}
\ref{Jensen, H.J. 1998, {\sl Self-Organized Criticality: 
	Emergent complex behaviour in physical and biological systems}, 
 	Cambridge University Press.}
\ref{Knizhnik, K.J., Uritsky, V.M., Klimchuk, J.A., and DeVore, C.R. 2018,
	ApJ 853, 82.}
\ref{Leka, K.D. and Barnes, G. 2003, ApJ 595, 1277.}
\ref{Lomax, K.S. 1954, J.~Am.~Stat.~Assoc. 49, 847.}
\ref{Lu, E.T., and Hamilton, R.J. 1991, ApJ 380, L89.}
\ref{Moon, Y.J., Choe, G.S., Park, Y.D., Wang, H., Gallagher, P.T.,
	Chae, J., Yun, H.S., and Goode, P.R. 2002, ApJ 574, 434.}
\ref{Pickands, J. 1975, Annals of Statistics 3, 119.}
\ref{Pruessner, G. 2012, {\sl Self-organised criticality. Theory, 
	models and characterisation}, Cambridge University Press, 
	Cambridge.}
\ref{Schrijver, C.J. 2007, ApJ 655, L117.}
\ref{Schrijver, C.J. 2009, AdvSpR 43, 739.}
\ref{Sharma, A.S. 2017, {\sl Data-driven modeling of extreme
	space weather}, in {\sl Extreme Events in Geospace}
	(ed. Buzulukova, N.), Elsevier.}
\ref{Sharma, A.S., Aschwanden, M.J., Crosby, N.B., Klimas, A.J., 
	Milovanov,A.V., Morales,L., Sanchez,R., and Uritsky,V.,
 	2016, SSRv 168, 167.}
\ref{Sheikh, M.A., Weaver, R.L., and Dahmen, K.A. 2016,
	Phys.Rev.Lett. 117/26, id. 26110.}
\ref{Sornette, D. 2004, {\sl Critical phenomena in natural sciences: 
	chaos, fractals, selforganization and disorder: concepts and tools},
 	Springer, Heidelberg.}
\ref{Sornette, D. 2009, J. Terraspace Science and Engeneering, 2(1), 1.}
\ref{Sornette, D. and Ouillon, G. 2012, in {\sl The European Physical
	Journal Special Topics  ``Discussion and Debate:
	From Black Swans to Dragon-Kings -- Is There Life Beyond
	Power Laws?}, 205, 1.}
\ref{Stumpf, M.P.H. and Porter, M.A. 2012, Science 335, 6069.}
\ref{T\"or\"ok, T., Panasenco, O., Titov, V.S., Mikic, Z.,
	Reeves, K.K., Velli, M., Linker, J.A., and De Toma G.
	2011, ApJ 739, L63.}
\ref{Uritsky, V.M., Davila, J.M., Ofman, L., and Coyner, A.J. 2013, ApJ 769, 62.}
\ref{Uritsky, V.M. and Davila, J.M. 2012, ApJ 748, 60.}
\ref{Watkins, N.W., Pruessner, G., Chapman, S.C., Crosby, N.B., and
	Jensen, H.J. 2016, SSRv 198, 3.}
\clearpage


\begin{table}
\setlength{\tabcolsep}{0.05in}
\tabletypesize{\footnotesize}
\caption{Results of forward-fitting of differential and cumulative size distributions for different 
statistical events, characterized by 
background counts (BG), 
the number of all events (N),
the number of events above threshold (NT),
the inertial range $\log(x_2/x_1)$ (IR),
the threshold value ($x_0$),
the power law slope from fitting a differential size distribution $\alpha_{diff}$, 
the power law slope from fitting a cumulative size distribution $\alpha_{cum}$, 
the goodness-of-fit values for both types of fits ($\chi_{diff}, \chi_{cum}$),
restricted to the range of the 10 most extreme events ($\chi_{ee}$), and 
the excessive extreme event factor ($q_{ee}$).}
\medskip
\begin{tabular}{lrrrrrrrrrrr}
\hline
&
BG &
N  &
NT &
IR &
$x_0$ &
$\alpha_{diff}$ &
$\alpha_{cum}$ &
$\chi_{diff}$ &
$\chi_{cum}$ &
$\chi_{ee}$ &
$q_{ee}$ \\
\hline
\underbar{Solar flares:}  & & & & & & & & & & & \\
Peak HXRBS)	   & 58 & 10856 & 7579 & 4.0 &   19 & 1.75$\pm$0.02 & 1.73$\pm$0.02 &  0.96 &  1.46 &  0.56 &  1.03 \\ 
Peak HXRBS         & 41 &  6223 & 3979 & 3.8 &   31 & 1.69$\pm$0.03 & 1.70$\pm$0.02 &  1.18 &  1.86 &  1.12 &  1.09 \\ 
Peak BATSE         &  0 &  7234 & 4996 & 3.3 &  501 & 1.91$\pm$0.03 & 1.81$\pm$0.02 &  3.18 &  4.39 &  0.78 &  1.11 \\ 
Peak RHESSI        &  0 &  7996 & 4399 & 3.7 &   20 & 1.87$\pm$0.03 & 1.86$\pm$0.02 &  1.28 &  2.00 &  0.77 &  1.88 \\ 
Peak RHESSI (A16)  &  0 &   290 &  149 & 1.7 &  822 & 2.44$\pm$0.19 & 2.28$\pm$0.13 &  0.70 &  1.07 &  0.81 &  1.45 \\ 
Counts HXRBS       &  2 &  8129 & 3314 & 4.6 & 3100 & 1.72$\pm$0.03 & 1.66$\pm$0.02 &  1.54 &  2.68 &  0.65 &  1.40 \\ 
Counts BATSE       &  0 &  2748 & 2748 & 4.8 & 2000 & 1.48$\pm$0.03 & 1.50$\pm$0.03 &  1.67 &  5.20 &  2.94 &  1.07 \\
Counts RHESSI      &  0 & 11548 & 8081 & 4.9 & 4783 & 1.60$\pm$0.02 & 1.64$\pm$0.02 &  3.34 & 10.98 &  2.77 &  1.54 \\ 
Counts RHESSI (A16)&  0 &   289 &  147 & 1.8 & $1.42 \times 10^6$ & 2.40$\pm$0.19 & 2.00$\pm$0.12 &  1.17 &  1.95 &  1.95 &  1.63 \\ 
Duration HXRBS     &  2 & 11539 & 4341 & 2.2 &  126 & 2.61$\pm$0.04 & 2.66$\pm$0.02 &  4.43 &  5.45 &  1.58 &  0.90 \\ 
Duration BATSE     &  0 &  7241 & 4562 & 2.1 &   37 & 2.32$\pm$0.03 & 2.19$\pm$0.03 &  5.65 &  5.81 &  0.00 &  1.77 \\
Duration RHESSI    &  0 & 11445 & 8296 & 2.2 &   25 & 2.05$\pm$0.02 & 1.90$\pm$0.02 &  7.59 &  8.28 & 10.19 &  0.93 \\ 
Duration (A14)     &  0 &   171 &  123 & 1.3 &    0 & 2.76$\pm$0.24 & 2.73$\pm$0.21 &  0.58 &  0.83 &  0.56 &  1.15 \\ 
Duration (A16)     &  0 &   289 &  189 & 1.0 &    0 & 3.22$\pm$0.22 & 3.33$\pm$0.20 &  1.53 &  1.49 &  0.64 &  1.83 \\ 
Length (A14)   	   &  0 &   171 &  112 & 0.9 &   26 & 5.13$\pm$0.44 & 5.03$\pm$0.38 &  1.10 &  0.52 &  0.55 &  1.25 \\ 
Length (A15)       &  0 &   389 &  210 & 0.8 &    8 & 4.23$\pm$0.27 & 4.45$\pm$0.23 &  1.18 &  1.00 &  1.04 &  0.95 \\ 
Area (A14)  	   &  0 &   171 &  100 & 1.5 &  899 & 2.84$\pm$0.27 & 2.82$\pm$0.22 &  0.71 &  1.05 &  0.73 &  1.75 \\ 
Volume (A14)  	   &  0 &   171 &   96 & 2.2 &30300 & 2.32$\pm$0.22 & 2.09$\pm$0.16 &  0.94 &  1.34 &  0.68 &  3.08 \\
Volume (A15)       &  0 &   389 &  239 & 2.1 &  640 & 1.83$\pm$0.12 & 1.86$\pm$0.09 &  0.56 &  0.72 &  0.55 &  1.07 \\ 
Magnetic energy (A14)&0 &   171 &   68 & 1.1 &  119 & 3.04$\pm$0.34 & 2.83$\pm$0.22 &  1.11 &  0.92 &  0.65 &  0.92 \\ 
Nonthermal energy (A16)&0&  192 &   89 & 2.6 &  0.32& 2.50$\pm$0.25 & 1.59$\pm$0.11 &  1.28 &  2.34 &  2.08 &  1.87 \\
Thermal energy (A15)& 0 &   390 &  279 & 1.8 &    3 & 2.15$\pm$0.13 & 2.16$\pm$0.11 &  0.86 &  0.73 &  0.73 &  1.02 \\ 
Emission measure (A15)&0&   391 &  251 & 1.4 & $6.5 \times 10^8$ & 2.97$\pm$0.19 & 2.82$\pm$0.14 &  0.86 &  0.58 &  0.49 &  1.00 \\ 
\hline
\underbar{Stellar flares:} &  & & & & & & & & & & \\
Peak KEPLER        &  0 &   208 &  124 & 2.8 &60255 & 1.66$\pm$0.14 & 1.68$\pm$0.12 &  0.81 &  1.26 & 1.45 & 3.55 \\ 
Bolom. KEPLER      &  0 &  1537 &  727 & 1.5 & $3.8 \times 10^{34}$ & 2.56$\pm$0.09 & 
                                                                      2.55$\pm$0.07 &  1.31 &  2.18 & 0.67 & 0.97 \\ 
\hline
\underbar{Terrestrial data:} &  & & & & & & & & & & \\
Earthquakes        & 1 &  17425 &  9423 & 4.6 &   100 & 1.80$\pm$0.02 & 1.85$\pm$0.01 &  2.22 &  3.11 &   1.45 & 2.82 \\ 
Fires              & 1 &  55853 & 38190 & 5.3 &     2 & 1.56$\pm$0.01 & 1.53$\pm$0.01 & 13.90 & 30.65 &  30.93 & 0.92 \\ 
Cities             & 0 &  19445 & 12784 & 4.0 &   501 & 1.79$\pm$0.02 & 1.79$\pm$0.01 &  5.69 & 18.37 &  14.18 & 1.79 \\ 
Blackouts          & 0 &    209 &   134 & 1.7 & 50000 & 1.98$\pm$0.16 & 1.90$\pm$0.13 &  0.80 &  0.83 & 0.63 & 3.35 \\ 
Terrorism          & 1 &   2699 &  1949 & 2.4 &     2 & 2.48$\pm$0.05 & 2.28$\pm$0.04 &  4.86 &  2.25 & 0.43 & 6.15 \\ 
Words              & 1 &   6609 &  4980 & 3.6 &     1 & 2.15$\pm$0.03 & 1.99$\pm$0.02 &  6.31 &  3.33 & 1.09 & 1.99 \\ 
Surnames           & 2 &   2280 &  1663 & 2.2 & 15700 & 2.55$\pm$0.06 & 2.50$\pm$0.05 &  1.10 &  1.89 & 1.34 & 1.13 \\ 
Weblinks           & 1 & $9.4 \times 10^7$ & $5.9 \times 10^7$ & 2.4 &  4 & 2.47$\pm$0.03 & 2.39$\pm$0.01 &  4.39 & 223 & 0.67 & 1.07 \\ 
\hline
\end{tabular}
\end{table}
\clearpage


\begin{table}
\tabletypesize{\normalsize}
\setlength{\tabcolsep}{0.05in}
\tabletypesize{\footnotesize}
\caption{Theroetically predicted power law slopes of solar flare size 
distributions (second and third column), which depend on the Euclidean 
dimension $(S)$, the fractal dimension $(D_S)$, the diffusion power 
exponent $(\beta)$, and the energy-volume scaling exponent $(\gamma)$, 
predicted by the fractal-diffusive SOC model (Aschwanden 2013). 
The power law slopes fitted to observed data sets (HXRBS, BATSE, RHESSI)
are listed in the fourth column. Measurements with poor statistics and
less reliable values are marked with parenthesis (...), obtained from
the data sets A14, A15, and A16 (Aschwanden et al.~2014, 1015, 2016).}
\medskip
\begin{tabular}{lllll}
\hline
Parameter		&Power law exponent	&Porwer law exponent 	      &Power law exponent      & Data set\\
			&(general expression)	&$S=3, \beta=1, \gamma=1$,	      &of observed solar flares& \\
			&			&$D_2=3/2, D_3=2$ 	      &$\alpha_x$              & \\
\hline
\hline
Length $L$              &$\alpha_L=S$		&$\alpha_L=3.00$	      &$\alpha_L=(4.45\pm 0.23)$ & A15    \\ 
Length $L$              &$\alpha_L=S$		&$\alpha_L=3.00$	      &$\alpha_L=(5.03\pm 0.39)$ & A14    \\ 
\hline
Area $A$                &$\alpha_A=1+(S-1)/D_2$	       &$\alpha_A=2.33$       &$\alpha_A=(2.82\pm 0.22)$ & A14    \\ 
Magnetic energy $E_{mag}$&$\alpha_{mag}\approx\alpha_A $&$\alpha_A=2.33$      &$\alpha_{mag}=(2.83\pm0.22$ & A14,A16\\
Emission measure $EM$	&$\alpha_{EM}\approx\alpha_A$  &$\alpha_A=2.33$       &$\alpha_{EM}=(2.82\pm0.14)$ & A15  \\
Bolom. energy $E_{bol}$ &$\alpha_{bol}\approx\alpha_A$ &$\alpha_A=2.33$       &$\alpha_{bol}=2.55\pm 0.07$ & KEPLER  \\ 
\hline
Volume $V$              &$\alpha_V=1+(S-1)/D_3$	&$\alpha_V=2.00$              &$\alpha_V=(2.09\pm 0.16)$ & A14    \\ 
Volume $V$              &$\alpha_V=1+(S-1)/D_3$	&$\alpha_V=2.00$              &$\alpha_V=(1.86\pm 0.09)$ & A15    \\ 
Thermal energy $E_{th}$	&$\alpha_{E_{th}}\approx\alpha_V$ &$\alpha_V=2.00$    &$\alpha_{th}=(2.16\pm0.11)$ & A15,A16 \\
\hline
Time duration $T$	&$\alpha_T=1+(S-1)\beta/2$ &$\alpha_T=2.00$ 	      &$\alpha_T=2.66\pm 0.03$ & HXRBS    \\
Time duration $T$	&$\alpha_T=1+(S-1)\beta/2$ &$\alpha_T=2.00$	      &$\alpha_T=2.19\pm 0.03$ & BATSE    \\
Time duration $T$	&$\alpha_T=1+(S-1)\beta/2$ &$\alpha_T=2.00$	      &$\alpha_T=1.90\pm 0.02$ & RHESSI   \\
Time duration $T$	&$\alpha_T=1+(S-1)\beta/2$ &$\alpha_T=2.00$	      &$\alpha_T=(2.73\pm 0.21)$ & A14,A16\\
\hline
Peak flux $P$		&$\alpha_P=1+(S-1)/(\gamma S)$ &$\alpha_P=1.67$ &$\alpha_P=1.70\pm0.02$ & HXRBS   \\
Peak flux $P$		&$\alpha_P=1+(S-1)/(\gamma S)$ &$\alpha_P=1.67$ &$\alpha_P=1.81\pm0.02$ & BATSE   \\
Peak flux $P$		&$\alpha_P=1+(S-1)/(\gamma S)$ &$\alpha_P=1.67$ &$\alpha_P=1.86\pm0.02$ & RHESSI  \\
Peak flux $P$		&$\alpha_P=1+(S-1)/(\gamma S)$ &$\alpha_P=1.67$ &$\alpha_P=(2.28\pm0.13)$ & A16   \\
Peak flux $P$		&$\alpha_P=1+(S-1)/(\gamma S)$ &$\alpha_P=1.67$ &$\alpha_P=(1.68\pm0.12)$ & KEPLER  \\
\hline
Fluence $C$		&$\alpha_C=1+(S-1)/(\gamma D_S+2/\beta)$ & $\alpha_C=1.50$ &$\alpha_C=1.67\pm0.02$ & HXRBS  \\
Fluence $C$		&$\alpha_C=1+(S-1)/(\gamma D_S+2/\beta)$ & $\alpha_C=1.50$ &$\alpha_C=1.50\pm0.03$ & BATSE  \\
Fluence $C$		&$\alpha_C=1+(S-1)/(\gamma D_S+2/\beta)$ & $\alpha_C=1.50$ &$\alpha_C=1.64\pm0.02$ & RHESSI \\
Fluence $C$		&$\alpha_C=1+(S-1)/(\gamma D_S+2/\beta)$ & $\alpha_C=1.50$ &$\alpha_C=(2.00\pm0.12)$ & A16  \\
Nonthermal energy $E_{nth}$&$\alpha_{E_{nth}}\approx\alpha_C$    & $\alpha_C=1.50$ &$\alpha_{E_{nth}}=(1.59\pm0.12)$ & A16 \\
\hline
\end{tabular}
\end{table}
\clearpage


\begin{figure}
\centerline{\includegraphics[width=0.9\textwidth]{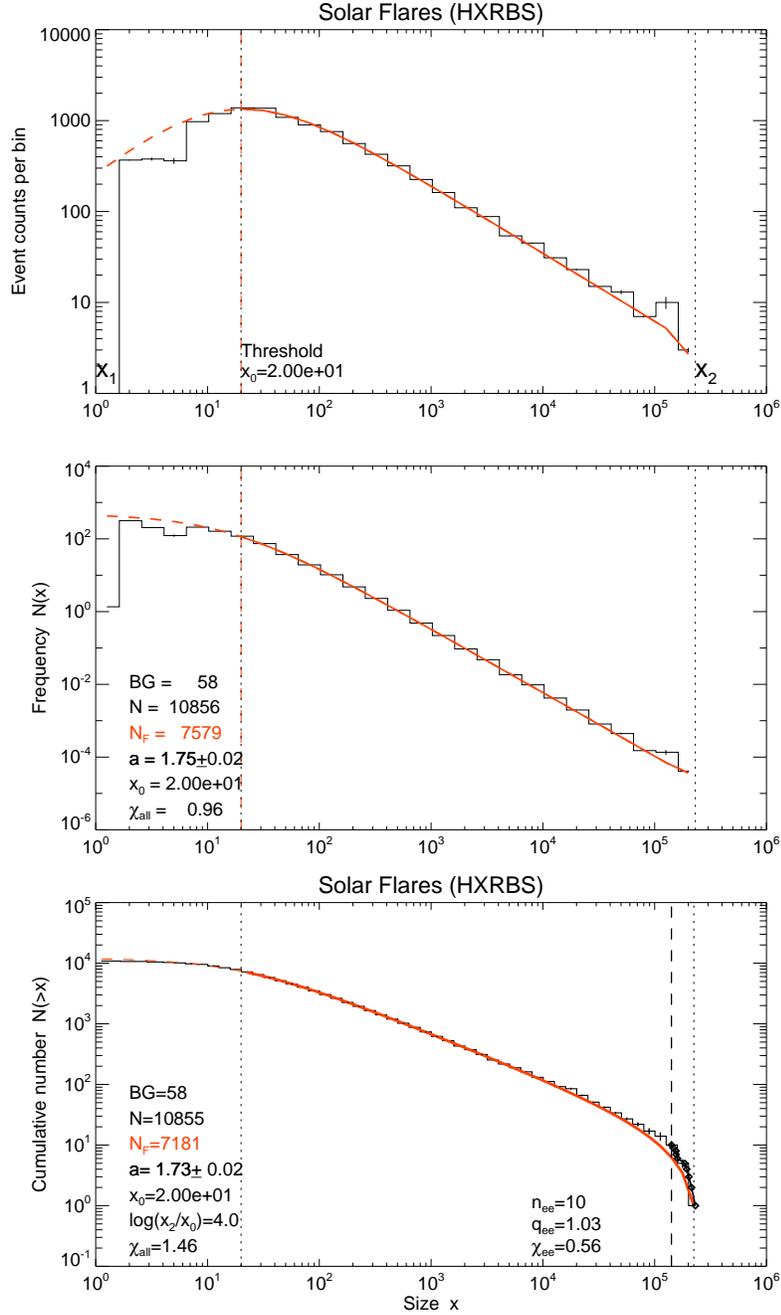}}
\caption{The statistics of solar flare hard X-ray peak photon rates
are given in form of a histogram with the number of events per logarithmic bin (a),
the differential occurrence frequency distribution (b), and the cumulative
occurrence frequency distribution (c), with the rank-order plot overlayed
for the 10 most extreme events (black diamonds). The observed data are shown in
form of black histograms, while the fitted thresholded power law function
is shown with a red curve. The data range is $[x_1, x_2]$ is marked with
dotted vertical lines, and the threshold value $[x_0]$ with a dotted dashed line.
The least-square fit criterion is calculated for the entire distribution 
sampled above the threshold ($\chi_{all}=1.46$), and for the range of
extreme events ($\chi_{ee}=0.56$).} 
\end{figure}

\begin{figure}
\centerline{\includegraphics[width=1.0\textwidth]{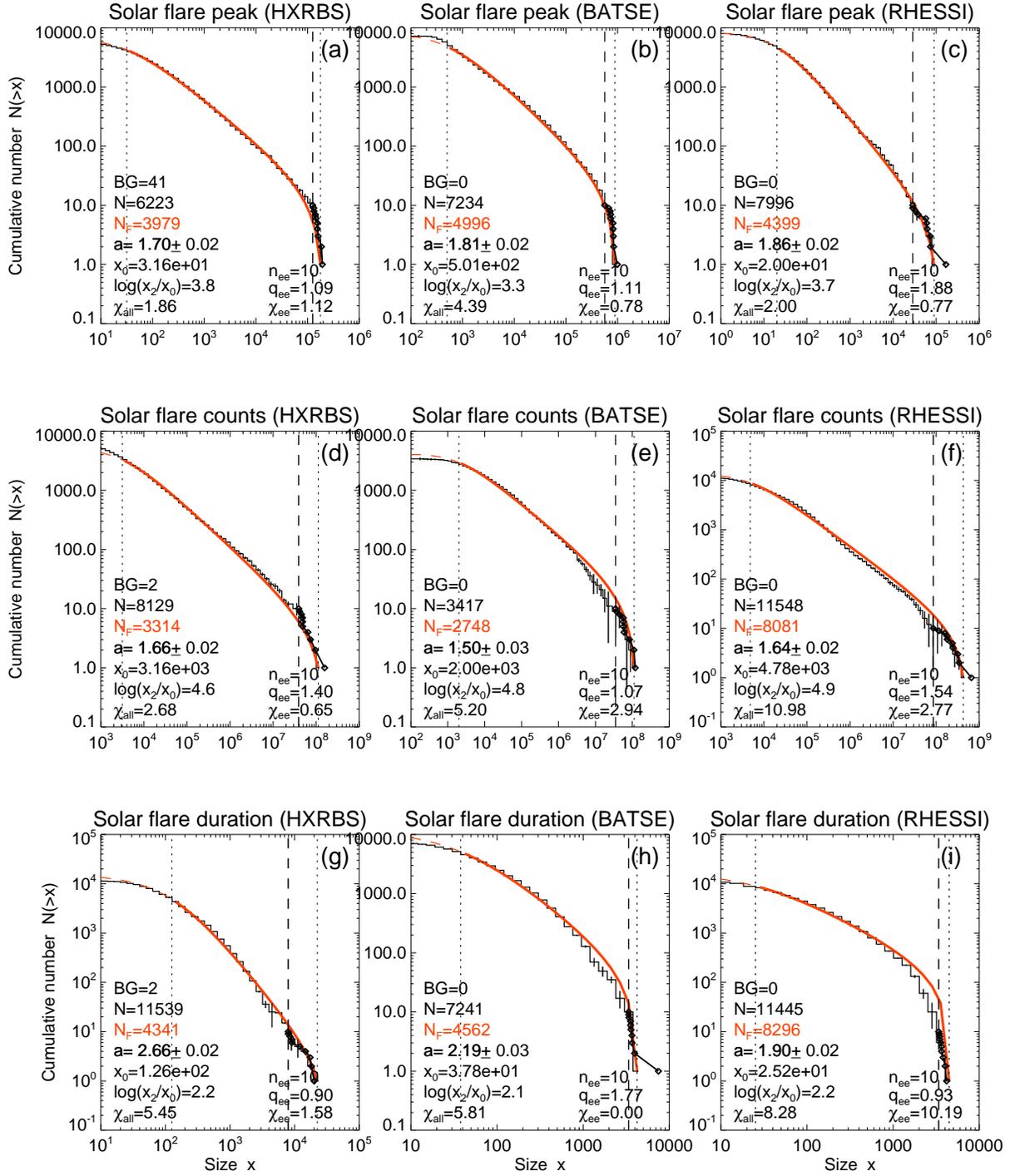}}
\caption{Solar flare hard X-ray peak photon rates (a-c), fluences or counts (d-f),
and flare durations (g-i) are histogrammed from HXRBS observations (a, d, g),
BATSE observations (b, e, h), and RHESSI observations (c, f, i) (shown in
form black histograms with uncertainties), and in form of the best-fit
cumulative distribution functions (red curves). The chi-square is given
for all events above the threshold $x_0$, $\chi_{all}(x > x_0)$, and for 
the 10 most extreme events separately, $\chi_{ee}(x > x_3)$.}
\end{figure}

\begin{figure}
\centerline{\includegraphics[width=1.0\textwidth]{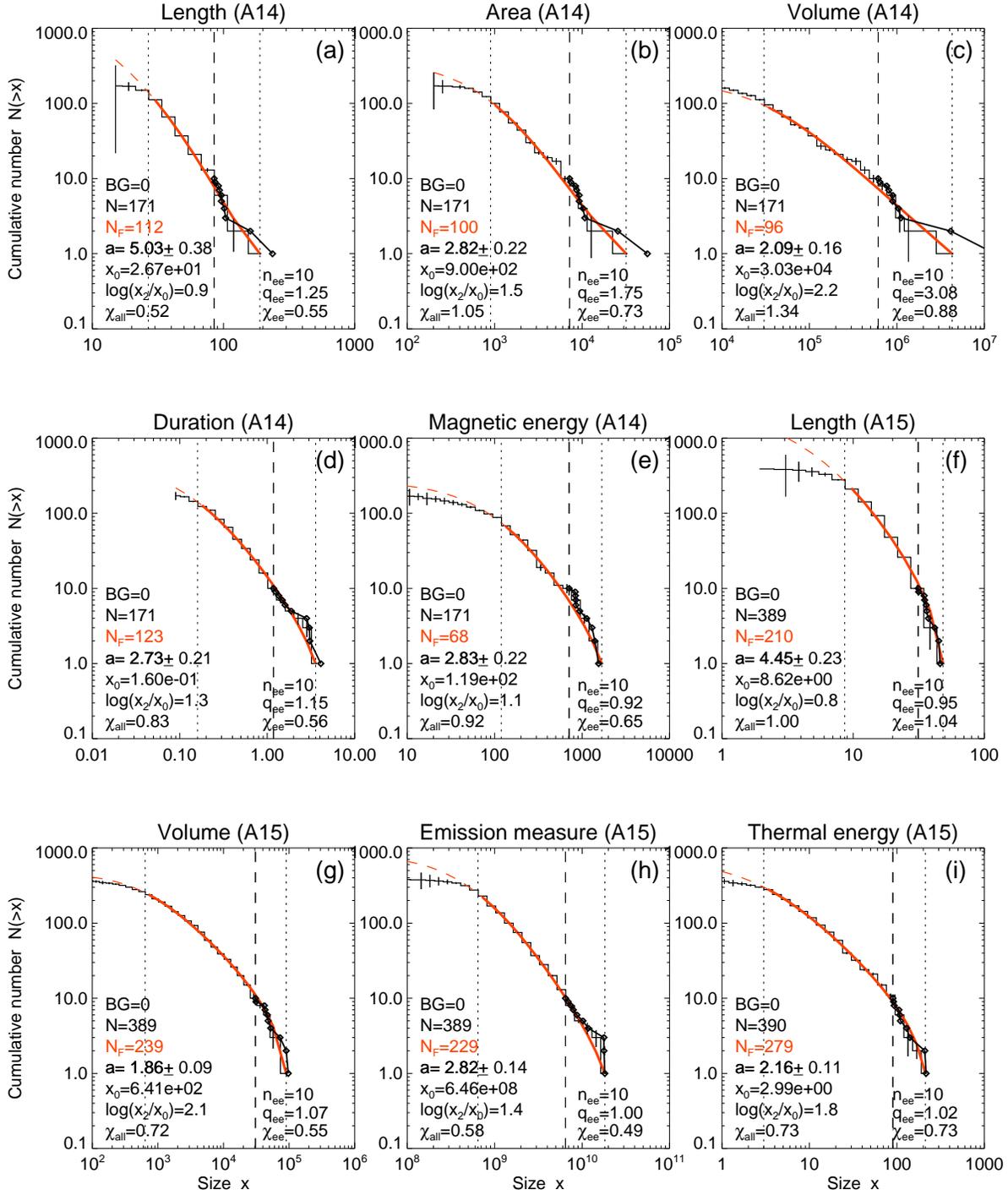}}
\caption{The cumulative size distributions of solar flare quantities 
are shown for length scales (a, f), flare areas (b) flare volumes
(c, g), flare durations (d), the dissipated magnetic energy (e),
the EUV emission measure (h), and the thermal energy (i),
calculated in two previous works (Aschwanden et al.~2014, 2015). 
Representation otherwise similar to Fig.~2).}
\end{figure}

\begin{figure}
\centerline{\includegraphics[width=1.0\textwidth]{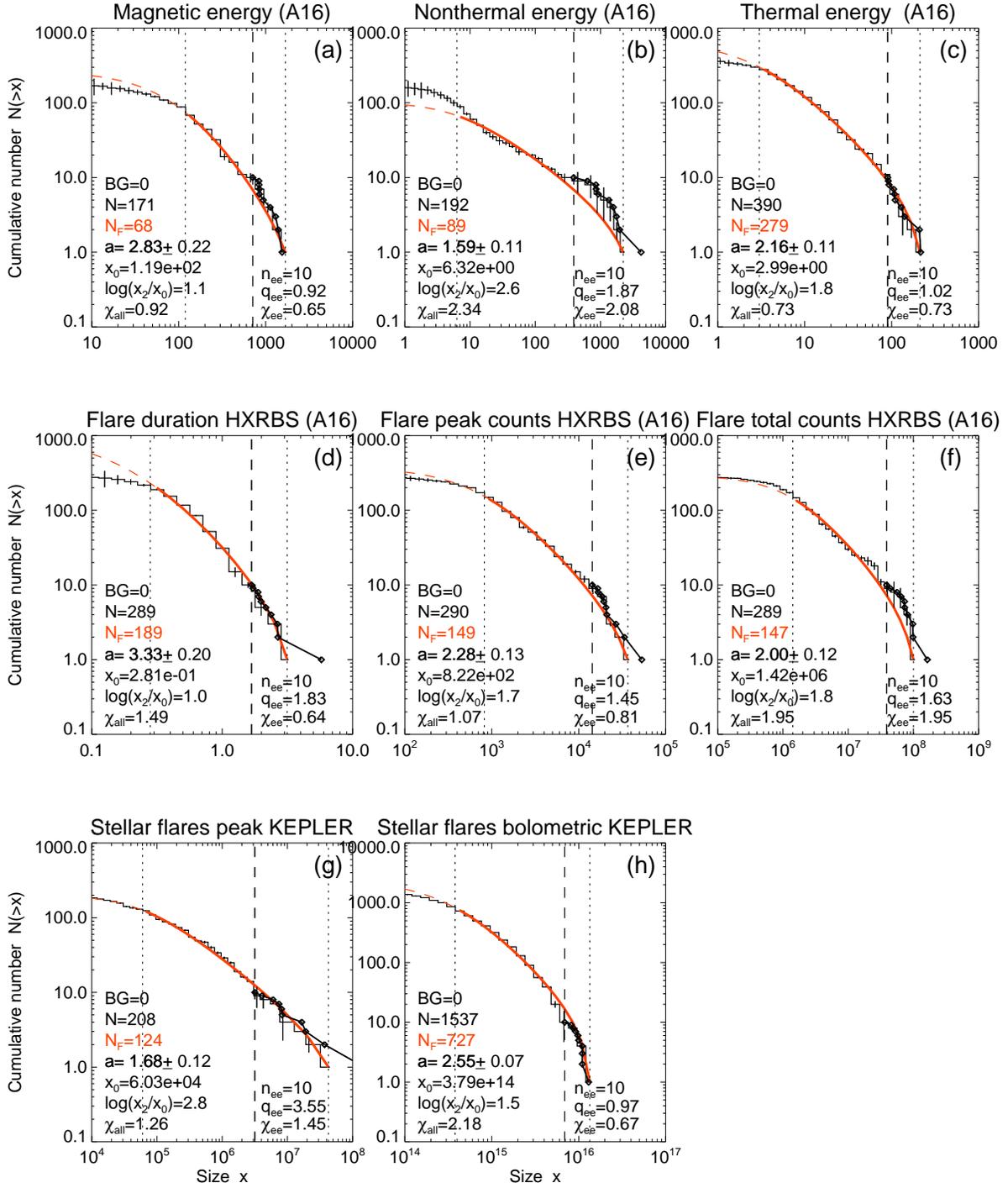}}
\caption{The cumulative size distributions of solar flare quantities 
are shown for the dissipated magnetic energy (a),
the nonthermal energy (b), the thermal energy (c),
flare durations (d), peak counts (e), total counts (f),
the peak intensity of stellar flares (g), 
and the stellar bolometric energy of stellar flares (h),
calculated from two previous works (Aschwanden et al.~2016). 
Representation otherwise similar to Fig.~2).}
\end{figure}

\begin{figure}
\centerline{\includegraphics[width=1.0\textwidth]{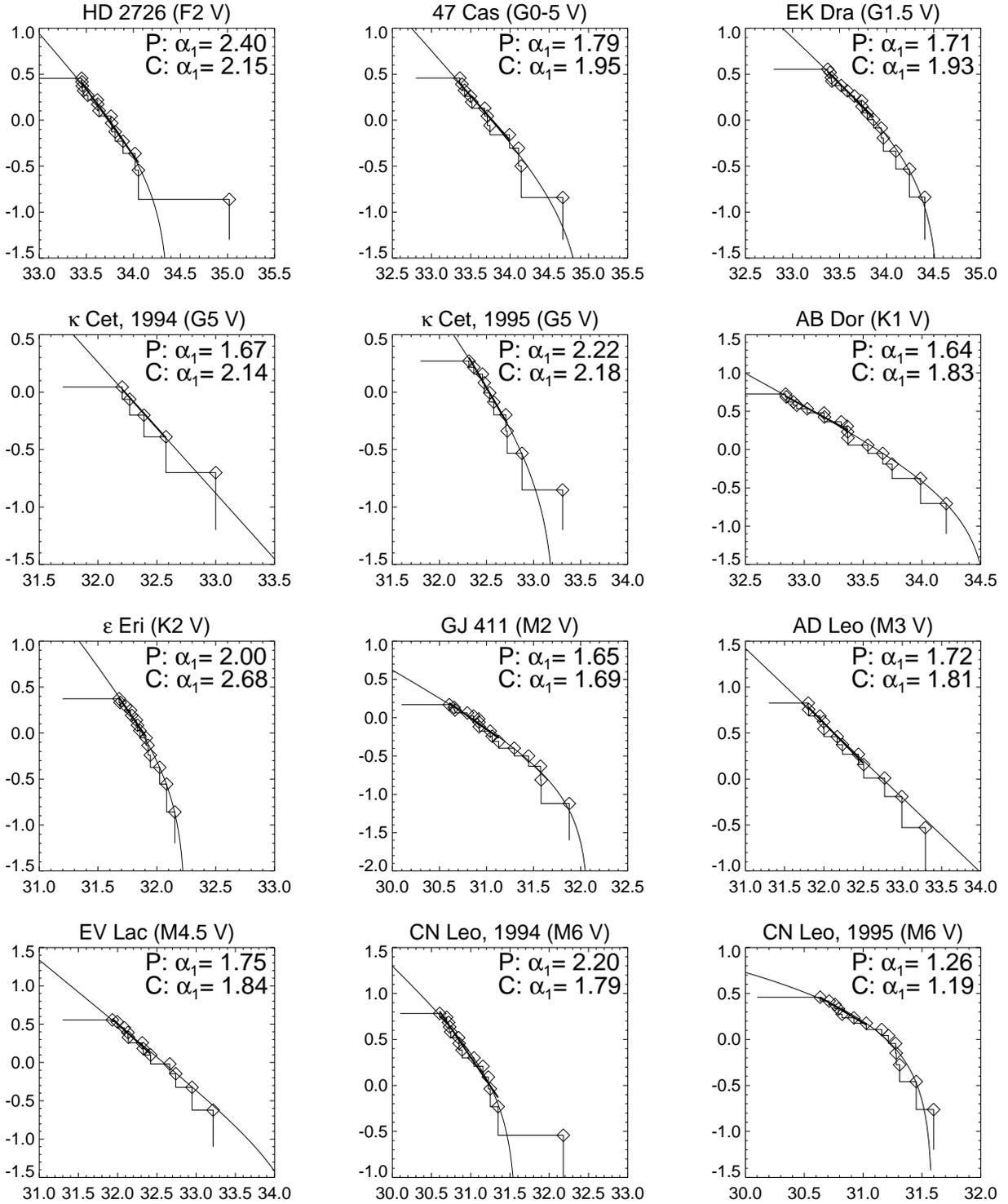}}
\caption{Cumulative frequency distributions of flare energies (total
counts) observed for 12 cool (type F to M) stars with EUVE
(Audard et al.~2000). The flare events are marked with diamonds,
fitted with a power law fit in the lower half (P; thick line), and
fitted with a cumulative frequency distribution (C; curved function).
Note two stars with outliers at the largest events (HD 2726) and
(CN Leo, 1994).}
\end{figure}

\begin{figure}
\centerline{\includegraphics[width=1.0\textwidth]{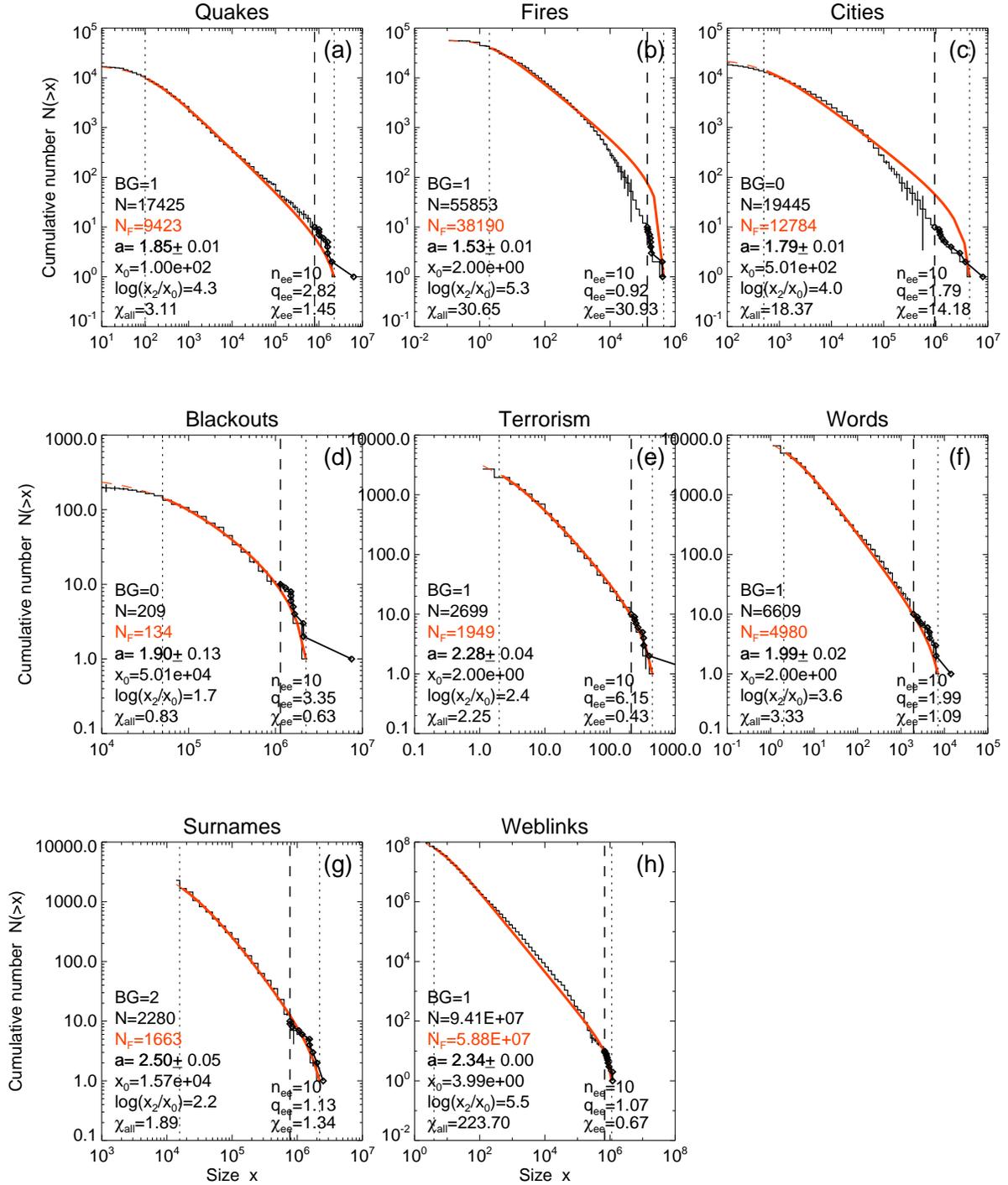}}
\caption{Cumulative size distributions are shown for 8 empirial data sets from
Clauset et al.~(2009): 
Earthquake intensities (a), forest fires (b), population of cities (c),
electric power blackouts (d), terrorist attack severity (e), count of
words (f), frequency of surnames (g), and links to web sites (h).
Representation otherwise similar to Figs.~2-4.} 
\end{figure}
\end{document}